\documentclass[preprint,aps]{revtex4}
\usepackage{graphicx}
\usepackage{mathptmx} 
\usepackage{natbib}

\begin{document}

\title{
Non-characteristic Half-lives in Radioactive Decay
}
\author
{
\'Alvaro Corral$^1$, Francesc Font$^1$, and Juan Camacho$^2$
}
\affiliation{
$^1$%
Centre de Recerca Matem\`atica, Edifici Cc, Campus Bellaterra,
E-08193 Barcelona, Spain
\\
$^2$%
Grup FESTA,
Facultat de Ci\`encies, Universitat Aut\`onoma de Barcelona, 
E-08193 Barcelona, Spain
}
\date{\today}

\begin{abstract}
{
Half-lives of radionuclides span more than 50 orders of magnitude.
We characterize the probability distribution of this broad-range data set
at the same time that explore a method for fitting power-laws and
testing goodness-of-fit.
It is found that the procedure proposed recently by Clauset {\it et al.}
[SIAM Rev. 51, 661 (2009)] does not perform well as it rejects the power-law hypothesis
even for power-law synthetic data.
In contrast, we establish the existence of a power-law exponent 
with a value around 1.1 for the half-life density,
which can be explained by the sharp relationship between decay rate and
released energy, for different disintegration types.
For the case of alpha emission, this relationship 
constitutes an original mechanism of power-law generation.
}
\end{abstract}

\pacs{
02.50.Tt,   
89.75.Da, 
23.90.+w 
}
\maketitle

\section{Introduction}

It is well known that most processes in physics are
governed by a characteristic scale.
Radioactive decay is a prototypical example of this;
for a given nuclide (i.e., an isotope of an element
in a given energy state), 
any nucleus has the same constant probability
of disintegration per unit time, 
which allows to define the half-life $t_{1/2}$ of the  
nuclide (or, equivalently, its lifetime), 
setting the time scale of the decay
\cite{Krane}.

It is noteworthy that these half-lives 
take very disparate values,
from very small fractions of a second
to many millions of years
(for example, 
$164$ $\mu$s for $^{214}$Po and
$4.47 \cdot 10^9$ years for $^{238}$U).
One can wonder if there is some predominant scale
for the half-lives of the radionuclides, 
or, on the contrary, half-lives can be considered as
scale free,
and which can be the physical reasons of that behavior.
It is remarkable that
in the last decades, many phenomena that violate 
the usual requirement of the existence of a characteristic
scale have been found in physics and beyond
\cite{Schroeder,Bak_book,Malamud_hazards,Sornette_critical_book,Christensen_Moloney},
under the form of power-law functions, 
the hallmark of scale invariance.
In parallel, the enormous range of values covered by the nuclear half-lives
will allow us to test the performance of a recently proposed statistical tool
to fit power-law distributions and 
to provide the goodness of such fits \cite{Clauset}.

\section{Data}

We analyze data coming 
from the Lund/LBNL Nuclear Data Search webpage \cite{Firestone}.
%
%
%
A total of 3032 nuclides are found from a
query in the range $10^{-34}$ s $<t_{1/2}<$ $10^{34}$ s
(to avoid 
249 ``stable'' nuclides as well as
433 ones with unknown half-lives, 
which are treated as having zero half-life in the web engine).
30 of these nuclides do not have a well-defined half-life,
as $t_{1/2}$ is only bounded from above or from below.
These, as well as the neutron
(which is not a radionuclide), 
are excluded from the analysis.
On the other hand, we include by hand the isotope $^{209}$Bi,
which was previously thought to be the heaviest stable nuclide
but has recently been found to be unstable, with $t_{1/2}=1.9 \cdot 10^{19}$ years
for $\alpha$ emission
\cite{Marcillac}.
We deal then with 3002 unstable nuclides, 723 of which are metastable
isomers (excited states with a ``relatively'' long half-life) and the rest,
2279, corresponding to ground states.
As we have found no big difference between the statistics
of the ground states and the isomers, we have considered both
types of radionuclides together in most of the analysis.
For a few of them (17) the half-life comes in units of energy, $E$,
which can be converted into half-life using the formula
$t_{1/2} E =\hbar$  
(where $\hbar$ is the reduced Planck's constant).
This yields values of $t_{1/2}$ below $10^{-16}$ s.

\begin{figure*}
\centering
\includegraphics[width=5in]{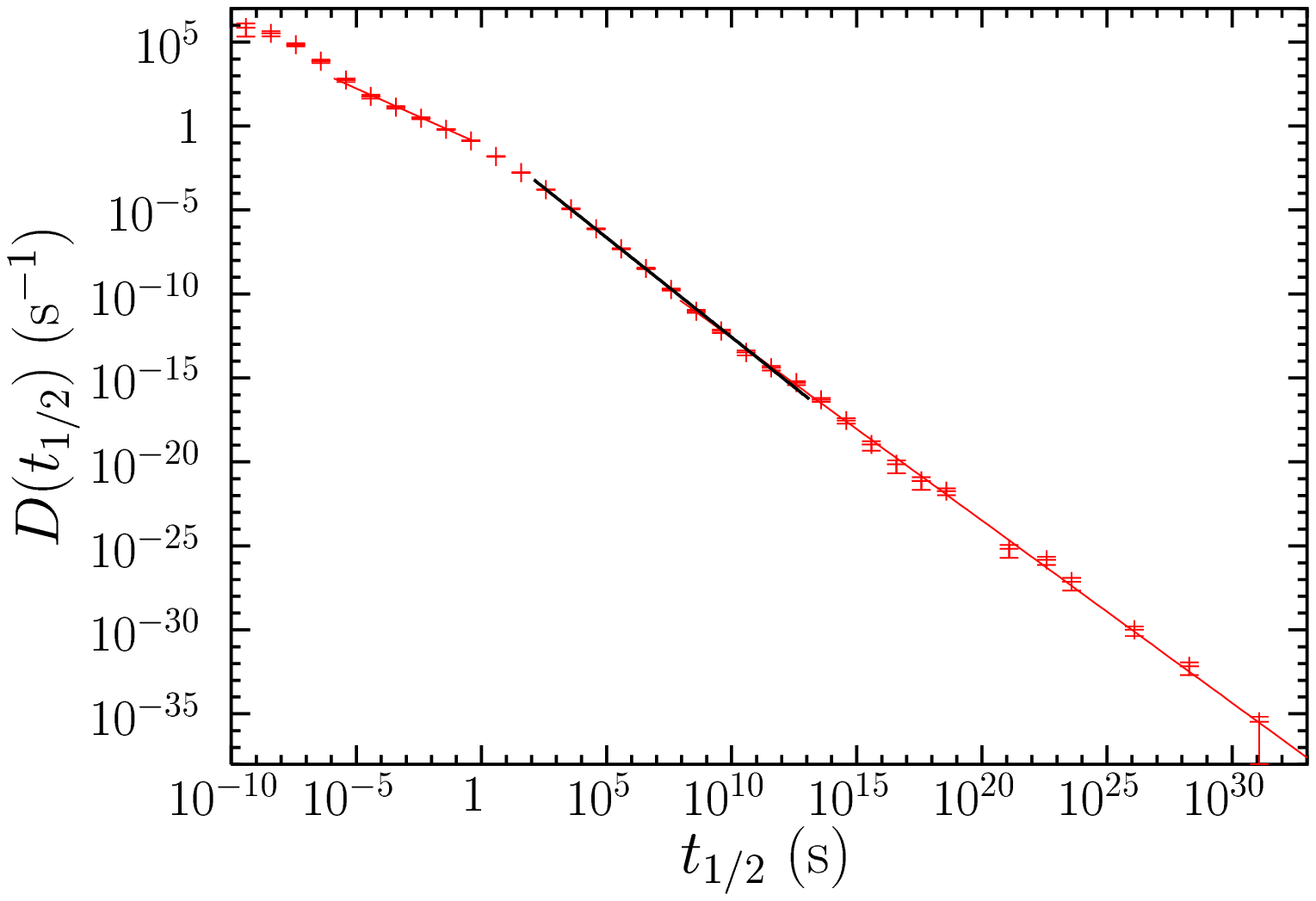}
\caption{
(Color online)
Empirical probability density of the half-lives of 
the 3002 radionuclides available 
(19 values below $10^{-10}$ s are not shown).
The straight lines are 3 of the power-law fits explained in the text,
covering the fitting range obtained by the minimization 
of the KS distance.
The exponents are $\tau=0.67$, $1.19$, and $1.09$ (from left to right).
Note that extreme caution should be present when estimating a probability density
using large bins, see Ref. \cite{Hergarten_book}.
\label{fig_density}
}
\end{figure*}

\section{Data Analysis}

A simple plot of the half-life probability density
on a log-log scale, displayed on Fig. \ref{fig_density}, shows that
the most prominent signature of the distribution is a linear trend for 
a certain range of half-lives,
an indication of a possible power-law behavior.
So, our first step is to test the existence of a power-law distribution
above a certain cutoff value $a$ of the half-life, i.e.,
$$
D(t_{}) \propto \frac 1 {t_{}^\tau} \mbox{ for }  t_{} \ge a,
$$
where $\propto$ denotes proportionality,
$D(t_{})$ is the probability density of the half-life, 
from now on denoted as $t_{}$,
and the exponent verifies $\tau > 1$ (for normalization).

We try the method proposed by Clauset {\it et al.} \cite{Clauset},
which looks for the value of $a$ which minimizes the Kolmogorov-Smirnov (KS) distance 
\cite{Press}
between the empirical cumulative distribution and the 
theoretical fit obtained by maximum-likelihood estimation of $\tau$
(values of $t_{}$ below $a$ do not play any role in the fit of $\tau$,
but are important for the estimation of $a$).
Once the values of $\tau$ and $a$ are found, a $p-$value is calculated, 
giving the probability that 
true power-law distributed data, with 
the same exponent and cutoff as the ones estimated for
the empirical data,
have a KS distance with its
fit larger than the distance of the empirical data
with its fit.
This is done by means of Monte Carlo simulations of the
resulting distribution, applying to the synthetic data
exactly the same fitting procedure 
as the one used for the empirical data
(maximum-likelihood estimation and minimization
of the KS distance), 
in order to avoid biases in the $p-$value
\cite{Clauset,Malgrem}.

The half-life distribution is easily simulated
in the power-law range (by means for instance of the inversion
method \cite{Press}), but it is necessary also to simulate it outside this range,
in order to optimize the value of $a$ in the synthetic data
(recall that the simulated distribution has to be treated in the same way
as the empirical one).
This is simply done by a kind of bootstrap method 
\cite{Clauset}, where the synthetic values of $t_{}$
are taken randomly, with replacement, from the empirical data
in that range (i.e., $t_{}<a$).
Details of the fitting and the goodness-of-fit testing
are explained in an appendix.
Noticeably, the whole fitting and testing procedure
does not use that $D(t_{})$ should be a straight line
in a log-log scale.

The results of this method applied to the nuclear half-lives
yield the values $a = 29.85 $ s 
and $\tau=1.16$
with a KS distance 
$d_m=0.036$ but with $p=0$ (from 1000 simulations,
that is, in no case a KS distance larger than the empirical 0.036 was found).
So, the direct application of the Clauset {\it et al.}'s method leads to the rejection
of the power-law hypothesis.

\begin{figure*}
\centering
\includegraphics[width=8.1cm]{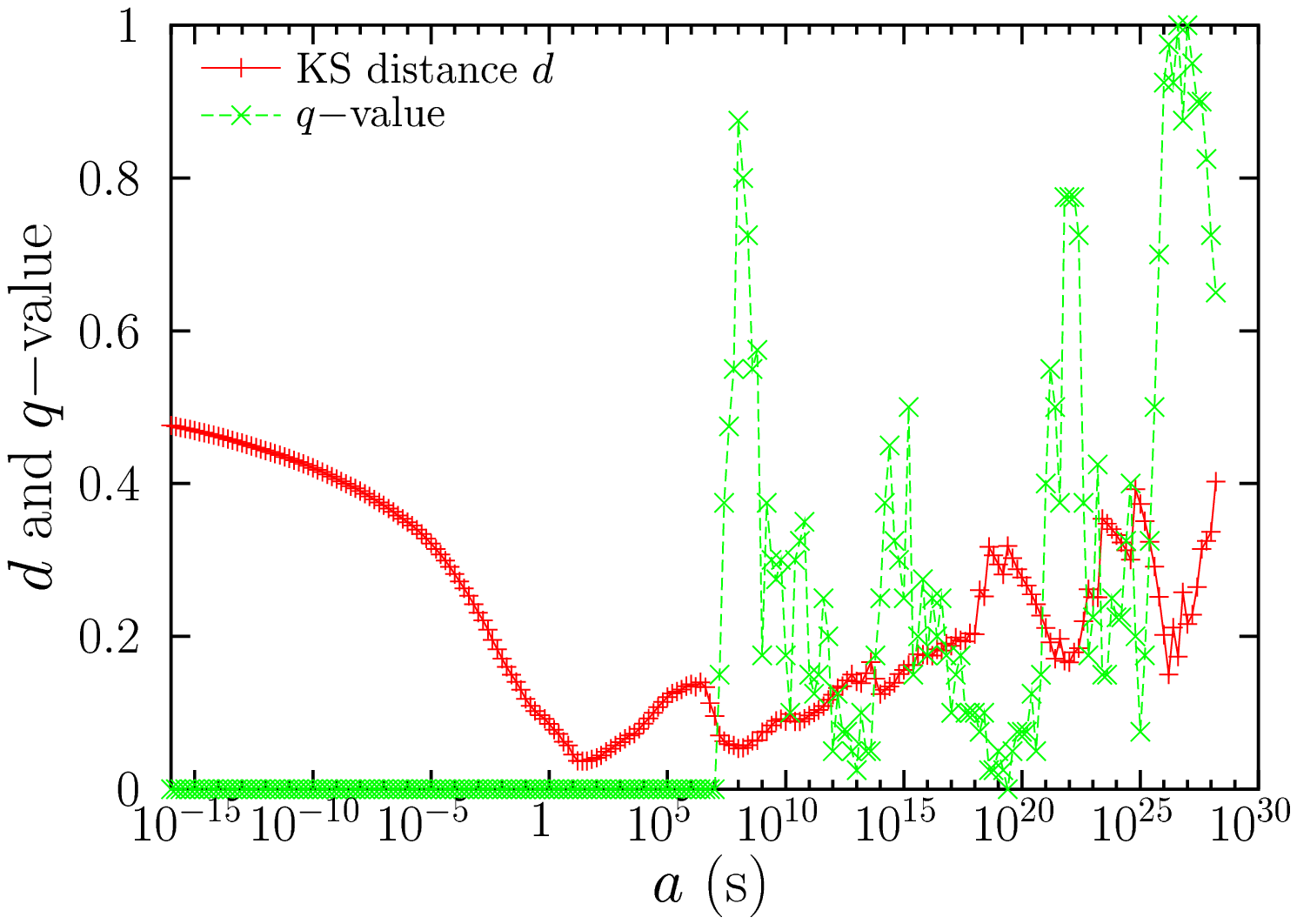}
\includegraphics[width=8.1cm]{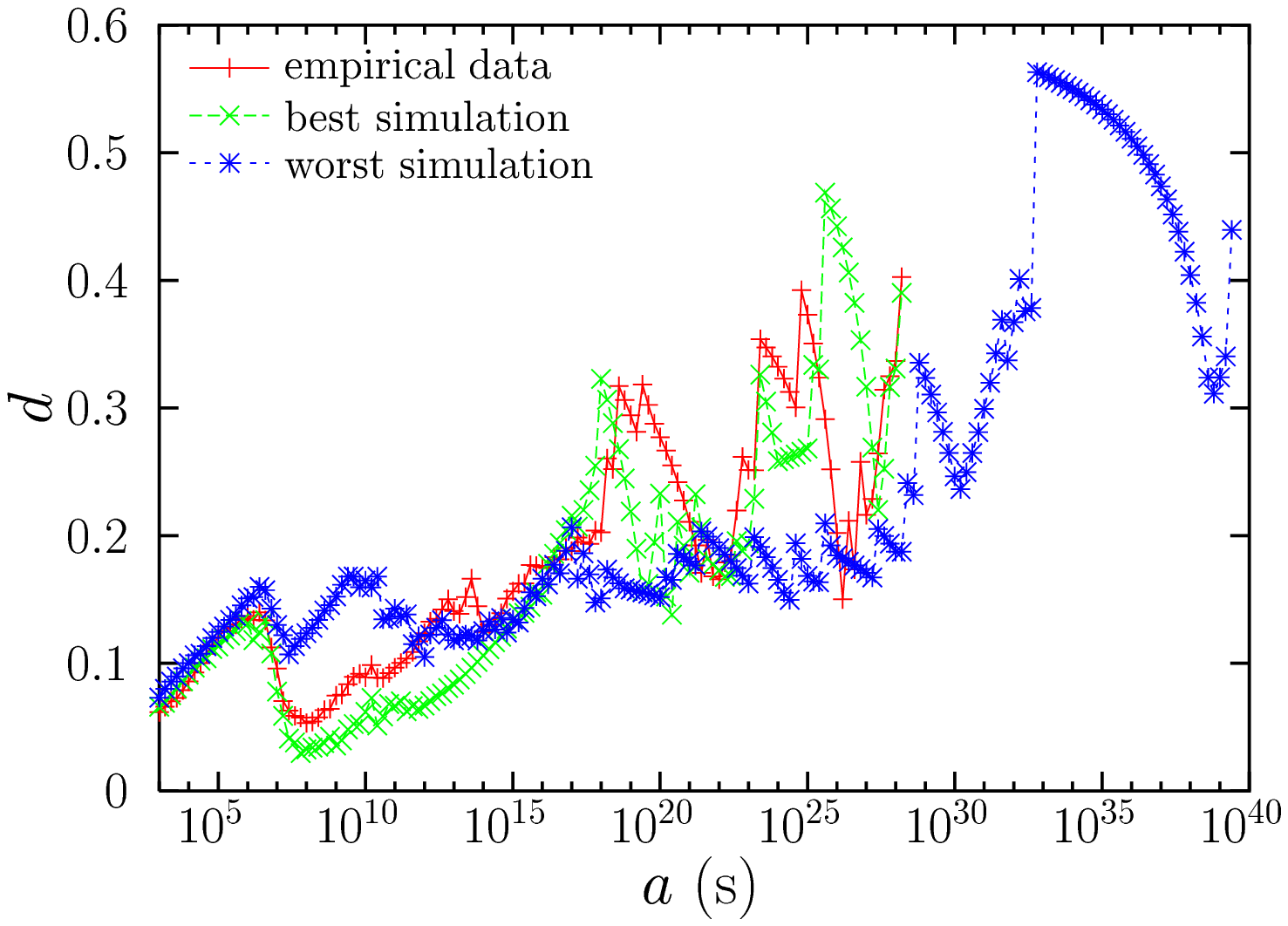}
\caption{
(Color online)
(a)
Kolmogorov-Smirnov distance $d$ for different values of 
the lower cutoff $a$ in a power-law fit
of the radionuclide half-life distribution,
together with the associated ``$q-$value'',
which would be identical to the $p-$value if the parameter
$a$ were not optimized. As $a$ is also variable in the fit, the $q-$value
overestimates the true $p-$value.
It is clear that the absolute minimum of $d$ yields $q=0$
and therefore $p=0$,
but for $a > 10^7$ s the $q-$value becomes nonzero.
(b) Comparison of the empirical KS distance in (a)
with that from simulations of a power law with exponent $\tau=1.09$
for $t_{} > 8.8  \cdot 10^7$ s
(and with random values taken from the empirical distribution below that
value, as explained in the text).
Only the simulations with the smallest ($d_m=0.028$, best) 
and largest KS distances ($0.073$, worst) out of 100 simulations are
shown. The fraction of minimum distances above the empirical value 0.052
defines the $p-$value.
\label{fig_test}
}
\end{figure*}

However, 
Fig. \ref{fig_test}(a) shows that this result is not convincing.
We  plot there, as a function of the cutoff $a$,
the KS distance and the $p-$value
corresponding to the case in which $a$ were fixed or known
(we denote the $p-$value for this case as $q$, in order to distinguish it
from the $p-$value when $a$ is optimized).
It is clear that the power-law fit must be rejected for any value of $a$
below $10^7$ s (as $q=0$),
but above this value $q$ takes non-zero values, fluctuating between zero and one,
as it would correspond to true power-law distributed data.

So, although the Clauset {\it et al.}'s method fails 
when applied to the whole data set,
we try to apply it now to a restricted data set, 
considering a range of variation of the parameter $a$ above 1000 s
(to avoid the misleading minimum of the KS distance at around 30 s).
This leads to a new minimum at $a = 8.8 \cdot 10^7$ s (close to 3 years)
and an exponent $\tau=1.09 \pm 0.01$
with $d_m=0.052$ and $p=33 \% \pm$ 5\%.
This is an acceptable result, which means in fact that
we cannot reject the power-law hypothesis.
Given that the maximum half-life is larger than $10^{31}$ s
(for $^{128}$Te),
this power law spans more than 23 orders of magnitude.
However, one can realize that there are only 128 nuclides in the power-law range, 
which makes its relevance as a characterization of nuclear properties rather limited.
In any case, we have found a power-law tail for the distribution
of nuclear half-lives,
together with the conclusion that the blind application of Clauset {\it et al.}'s
method is not reliable.
The condition $a >1000$ s is not determinant, 
as other values lead to very similar results,
see Table 1.


In order to confirm the failure of Clauset {\it et al.}'s
method we apply the absolute minimization of the KS distance 
to simulated data,
with a power-law distribution with $\tau=1.09$ above $a=8.8 \cdot 10^{7}$ s
and bootstrapping the empirical data below that value \cite{Clauset}
-- the KS distance corresponding to two of them is shown in Fig. \ref{fig_test}(b).
In 80 \% of the simulations the $p-$value turns out to be zero,
leading to the rejection of the power-law hypothesis
when the data have, by construction, a true power-law range.
Notice that in this case, as the null hypothesis is true, the $p-$value
should be uniformly distributed between 0 and 1.
Therefore, we have come across a counterexample that invalidates the general 
applicability of the Clauset {\it et al.}'s method.
There is nothing surprising here, as,
after all, these authors do not provide any theoretical 
justification about why their ad-hoc method should work, 
and only check its performance (under controlled conditions) in simple 
synthetic examples.

Our second option for the distribution of half-lives 
is the use of an upper truncated power-law distribution,
$$
D(t_{}) \propto \frac 1 {t_{}^\tau} \mbox{ for } a \le t_{} < b,
$$
where $a$ and $b$ are the lower and upper cutoffs, respectively,
with no condition on the value of $\tau$ this time.
Some examples of the applicability of this distribution can be
found in Ref. \cite{Burroughs_Tebbens}. 
In this case we need to generalize Clauset {\it et al.}'s method
(a detailed description is provided in the supplementary information of
Ref. \cite{Corral_hurricanes}, of which our appendix gives a summary).
The basic idea is again to find the values of $a$ and $b$ which minimize
the KS distance between the empirical distribution and its fit.
An important difference with the previous case 
is that when $b$ is not infinite, several power-law pieces can 
coexist in the data,
and a criterion is necessary to select one of them.
Restricting this time the $a-$value to $a > 1$ s,
the pair that minimizes the KS distance
(determined with a resolution of 5 points per decade) is
$a= 126$ s and $b=7.9 \cdot 10^6$ s,
with $\tau=1.18 \pm 0.01$,
$d_m=0.012$, and $p=88\% \pm 3\%$.
Therefore,
a power law ranging more than 4 decades cannot be rejected.

However, this solution, with its high $p-$value, is not fully satisfactory,
as its range is a small fraction of the total half-life range.
We have observed, by means of computer simulations,
that the generalization of the Clauset {\it et al.}'s 
method that we are using leads to severe underestimations of the value of $b$,
of many orders of magnitude in some realizations.
Adding a different restriction, for example $b/a > 10^{8}$,
we find the conditional minimum at $a=126$ s and $b=1.3 \cdot 10^{13}$ s,
with an exponent $\tau=1.19 \pm 0.01$, 
a KS distance $d_m=0.015$ and a $p-$value $p=39\% \pm 5\%$,
comprising 1277 different nuclides.
We conclude that a power law over 11 orders of magnitude
is an acceptable fit 
(although higher values of $b$ still yield not too low $p-$values).
All the fits are collected in Table 1
and the most representative ones are shown in 
Fig. \ref{fig_density} (see caption for details).
A more standard analysis based on the calculation of the $q-$values
(again, $p-$values where $a$ and $b$ are fixed,
as in Ref. \cite{Peters_Deluca})
show that indeed these power-law fits are acceptable.


The results found so far can be summarized
by two power-law ranges, 
one from, roughly, $t_{}=100$ to $10^{10}$ s, 
with $\tau=1.19$,
and another one for $t_{}$ above $10^{10}$ s,
with $\tau=1.09$ (the crossover region 
between the two exponents is difficult to discriminate).
The close value of the exponents 
lead us to speculate that the two regimes
could be in fact the same, with some artifact 
causing the small but significant difference
between their values.
After all, we have studied about 3000 nuclides,
but of course these do not constitute a complete sample.
Among the nuclides considered as stable, more than 100
are theoretically unstable, but with a half-life
that has not been possible to measure.
And of course, not all nuclides that may exist
have been synthesized so far \cite{Miller}.
So, we leave the door open to the fact that a unique
exponent around $\tau=1.15 \pm 0.10$ could describe
the complete set of radioactive half-lives above
the range of a few minutes.
This would imply the existence of scale invariance 
for about 30 orders of magnitude.

It is interesting to see how these results are 
affected by the mode of disintegration of the radionuclides.
Different decays involve different processes
and even different types of interaction
(strong for $\alpha$ emission and
weak for $\beta$, for instance).
We can restrict the half-life statistics to nuclides
that decay in a particular way.
Table 2 shows the results corresponding to the 
$\alpha$, $\beta$, electron-capture, and isomeric-transition decays,
including also the separation of all nuclides 
into isomers and ground states. 
To be unambiguous, when we consider $\alpha$ disintegration, 
only those radionuclides that decay exclusively 
by $\alpha$ emission are considered, 
and the same for the other types of disintegration.
We fit a double-truncated power-law distribution
concentrating on the region of large half-lives
($a>1$ s) and
see that in all cases the tail of the distribution
is compatible with a power law with an exponent 
between 1 and 1.3.

\section{Origin of Power-law Behavior}

In the case of $\alpha$ decay we can give a simple explanation 
of the exponent.
Gamow's model assumes the pre-existence of an $\alpha$ particle
inside a potential well \cite{Krane}. Quantum tunneling of the particle
leads to a very sharp relation between the half-life and the energy $Q$ released
in the emission
(which can be related to the Geiger-Nuttall rule),
$$
t_{}=A \exp\left(\frac{B Z}{\sqrt{Q}}\right),
$$
where $Z$ is the atomic number of the resulting nucleus, 
$B$ is a positive constant, and $A$ can be considered as a constant too,
in a first approximation \cite{Williams}.
Defining $U=Q/{Z^2}$, with an unknown probability density $f(U)$,
conservation of probability leads to
$$
D_\alpha(t_{})=f(U)\left|\frac {dU}{dt_{}}\right|=
f\left(\frac{ B^2}{\ln^2\,{t_{}}/ A}\right) 
 \left(\frac{2B^2}{\ln^3\,{t_{}}/ A}\right)\frac 1 {t_{}},
$$ 
where the factor $F(t)$ that multiplies the hyperbolic $1/t$ tail
constitutes a slowly varying function for 
a broad class of choices for $f(U)$
(i.e., $F(\ell t)/F(t) \rightarrow 1, \forall \ell >0 $ when $t\rightarrow \infty$).
For example, $f(U)$ could be uniform, linear, parabolic, hyperbolic,
etc., with no influence on the $1/t$ asymptotic tail.
This result is in surprising good agreement
with the empirical exponent found for the $\alpha$ decay,
$\tau=1.0$, see Table 2. 
Note that the present mechanism for power-law generation
has some resemblance with the ones collected in Refs. 
\cite{Sornette_critical_book,Newman_05},
but, in contrast to them, it is not derived neither from another 
power-law relationship between $t$ and $U$ (see next)
nor from an exponential energy-barrier distribution.

For the $\beta$ decay a similar argument is possible.
In this case, Sargent rule establishes a different relationship
between $t_{}$ and the released energy $Q$
\cite{Perkins},
which, in a very crude approximation, could be considered as 
a proportionality between $t_{}$ and $1/Q^5$, so, 
$$
Q \simeq \frac C {t_{}^{1/5}}
$$
Then, the distribution of $t_{}$ could be obtained
from that of $Q$, $g(Q)$,
$$
D_\beta(t_{}) \propto g\left(\frac C{t_{}^{1/5}}\right) \frac 1 {t_{}^{1+1/5}},
$$
which yields asymptotically a power law with exponent 1.2 if $g$ is a slowly varying
function of $t_{}$.

However, the fact that $C$ could be constant is far from true, 
rather, it varies in a range of many orders of magnitude 
for different nuclides \cite{Krane}.
A mathematical theorem guarantees that, under very general conditions, the 
values above a high threshold of a random variable tend
to follow the so-called generalized Pareto distribution \cite{Coles,Embrechts},
which asymptotically yields a simple power-law tail
when the range of variation is broad
(decay slower than an exponential).
So, it is reasonable to assume that, over some threshold value, 
the half-life is given by the product of 
two power-law variables, one with an exponent 6/5 (derived above from Sargent rule) 
and the other one unknown, $1+\nu$.
Considering independence between both factors, the distribution of the product
is given by the convolution of the logarithms of each factor,
which turn out to be exponentially distributed. 
The result is,
$$
D_\beta(t_{}) \simeq \frac {\nu/5}{ \nu-1/5} \left( \frac {K^{1/5}}{t^{6/5}}
- \frac {K^{\nu}}{t^{1+\nu}} \right),
$$
where $K$ is the minimum value of $t$
for which this behavior holds.
This yields a power-law tail with exponent $\tau=1+1/5$
if $\nu > 1/5$ and with exponent $\tau=1+\nu$ 
if $\nu < 1/5$ (but for normalization, $\nu >0$).
In any case, the exponent $\tau$ of the tail is constrained 
in the range $(1, 1.2)$.

We notice that the gamma decay also leads to power-law relations
between the half-life and the released energy \cite{Krane}, with even exponents
from $t_{} \propto 1/Q^3$, $t_{} \propto 1/Q^5$, 
etc.;
this means power-law distributions with $\tau\simeq 4/3$, 6/5...
very close to one in all cases,
in agreement with the results found for isomeric transitions.

Power-law behaviors as the ones found here for radioactive decays are widely spread in nature. 
The origin of the abundance of these scaling laws is a topic of great interest in our understanding of complex systems, 
i.e. systems containing a large number of highly interacting components. 
Nuclei are a good example of a complex system, as nucleons strongly interact among each other 
in such a 
way that no theory is able to predict from first 
principles which nuclides will be stable and which not. 
Though one might be tempted to believe that just a general mechanism 
may be responsible for the ubiquity of power laws in nature, 
several different explanations have been reported, such as aggregation processes 
\cite{Vicsek,Camacho_Sole}, intermittency \cite{Zeldovich}, 
self-organized criticality \cite{Bak_book}, and multiplicative processes 
\cite{Reed,Saichev_Sornette_Zipf},
see Refs. \cite{Sornette_critical_book,Mitz,Newman_05} for reviews. 
As discussed above, the power laws of nuclear half-lives
are simply related to the extremely steep relationships between
half-life and energy, which translate small changes
in energy to variations of orders of magnitude 
in half-life.

\section{Other Power Laws and Other Fits}

Going back to data analysis, 
one can perform the same study with the 
half-lives of elementary particles.
We have compiled a list of 131 baryons and mesons,
with $t_{}$ ranging from $10^{-25}$ s
to about 10 min (for the neutron).
The direct application of the Clauset {\it et al.}'s
method, described above, leads to a pure power-law
distribution with $a=2.4 \cdot 10^{-14}$ s
and $\tau=1.13 \pm 0.03$, comprising 19 particles and
with $p=7\% \pm 2.5 \%$.
Although this is in the limit of rejection, 
it is striking that the exponent $\tau$ is essentially
the same as for the radionuclides.
The similarity in some disintegration processes
could explain this behavior. 

Interestingly, other power-law ranges exist for the distribution
of nuclear half-lives.
For the whole dataset, imposing $b< 1$ s
(which corresponds to the region
of short half-lives), we find the optimum values
$a=5.0 \cdot 10^{-5}$ s and $b=0.32$ s,
with $\tau=0.65 \pm 0.02$, $d_m=0.023$ and $p=40\% \pm 5\%$,
comprising 460 radionuclides.
We can conclude that a second (or a third), flatter power-law
regime is present in the data, 
although incompleteness of the data for short half-lives
can affect the range of this regime, see Table 1.

The transition between the short and long half-life
power-law regimes shows a convex shape in log-log scale
that is well modeled by a (truncated) lognormal 
distribution from about $10^{-2}$ to $10^{5}$ s,
with the parameters of the underlying (untruncated) normal
distribution ($x\equiv \ln t_{}$) given by $\mu\simeq 3.4$ and
$\sigma^2\simeq 30$, if $t_{}$ is measured in seconds.
Again, there is an overlap region between the different distributions, 
in which a power-law fit and the log-normal fit are both valid.

We are not interested in determining precisely when 
one distribution transforms into another, 
but, more importantly, we have tested
that the lower truncated lognormal distribution
is not preferred in front of the power law
for values of $t_{}$ above 100 s.
This has been done by the straightforward adaptation
of the uniformly most powerful unbiased test proposed in Ref. \cite{Castillo}. 
The main idea is that a power-law distribution
constitutes a special case of a (truncated) lognormal distribution,
achieved when $\sigma \rightarrow \infty$ and $\mu \rightarrow -\infty$
but with $\sigma \ll |\mu|$ in such a way that the power-law
exponent turns out to be $\tau=1+|\mu|/\sigma^2$.
A likelihood ratio test evaluates if other parameters for the lognormal
are preferred or not. In our case the power-law fit cannot be rejected.
The non-preference of the lognormal can be extended
to other related distributions \cite{Castillo}.






\section{Conclusion}

In summary, the distribution of half-lives
of the radionuclides has a power-law tail, 
with an exponent slightly greater than 1,
due to the abrupt relationship between 
decay rate and released energy for the
different transitions analyzed.
The broad range of scales covered by the distribution
and the scarce number of nuclides in the 
right-most part of it makes difficult
the use of standard statistical tools, 
and even a recently introduced fitting
procedure \cite{Clauset} fails to detect the power law.
Careful analysis is necessary when dealing with power-law distributions
with such small exponents.

\section{Acknowledgements}

The authors acknowledge valuable information flow from
A. Clauset, A. Deluca, R. D. Malmgren, P. Puig, and K. Sneppen,
and appreciate the efforts of the late P. Bak, B. B. Mandelbrot,
and M. Schroeder to point out the relevance of power-law statistics.
F. F. enjoyed a grant of the CRM for the realization of
his master's thesis. Spanish 
projects involved: 
FIS2009-09508, FIS2009-13370-C02-01 and 2009SGR-164. 
A.C. also participates in the Consolider i-Math project.


\section{Appendix}

Our fitting procedure and the testing of the goodness of fit
follow closely the ideas reported by Clauset {\it et al.}
\cite{Clauset}, and are explained in more detail in the supplementary 
information of Ref. \cite{Corral_hurricanes}.
Here we just provide and overview of the basic method,
further modifications are specified in the main text.

The maximum-likelihood (ML) estimation of the exponent of a power-law distribution
of the form $D(t) = (\tau-1) a^{\tau-1}/t^\tau$, with $t>a$ (and $a$ known) is given by
$$
\tau = 1 +\frac 1 {\ln \frac {G_a} a},
$$
where 
$G_a$ is the geometric mean of the data above the cutoff,
$\ln G_a \equiv N_a^{-1}\sum_{t \ge a} \ln t$,
and $N_a$ is the number of such data.
It can be easily shown that for true power-law distributed data
the ML estimator of the exponent (minus 1) follows an inverse gamma distribution, 
with mean $N_a (\tau-1) /(N_a-1)$ 
asymptotically equal to the true exponent (minus 1) and a standard deviation
given by 
$$
\varepsilon=\frac{\tau-1}{(1-N_a^{-1})\sqrt{N_a-2}} \rightarrow 
\frac{\tau-1}{\sqrt{N_a}};
$$
note that the uncertainty of the exponent
goes to zero as its value goes to one
(which on the other hand is forbidden by normalization).

\begin{figure*}
\centering
\includegraphics[width=8.1cm]{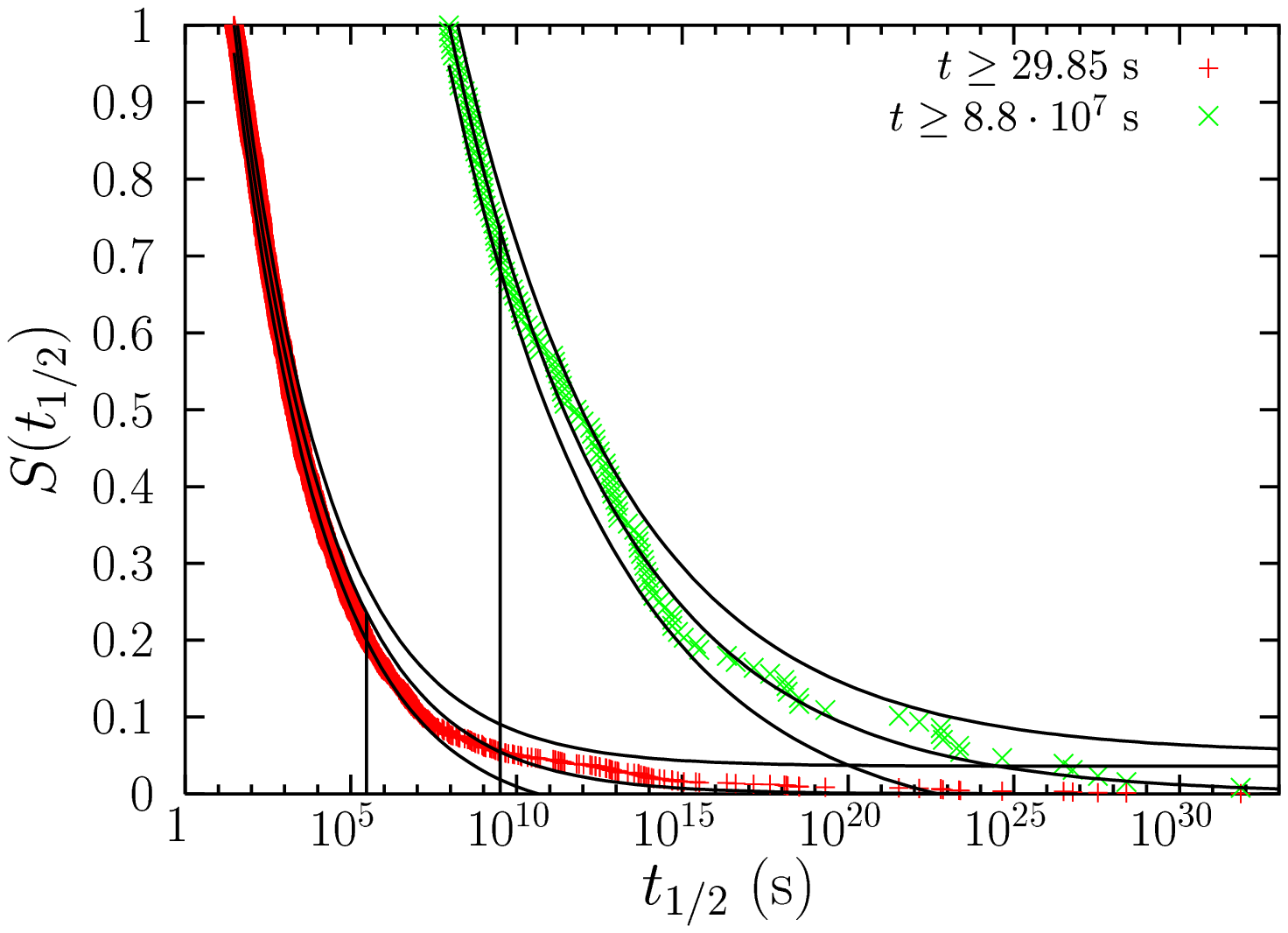}
\includegraphics[width=8.1cm]{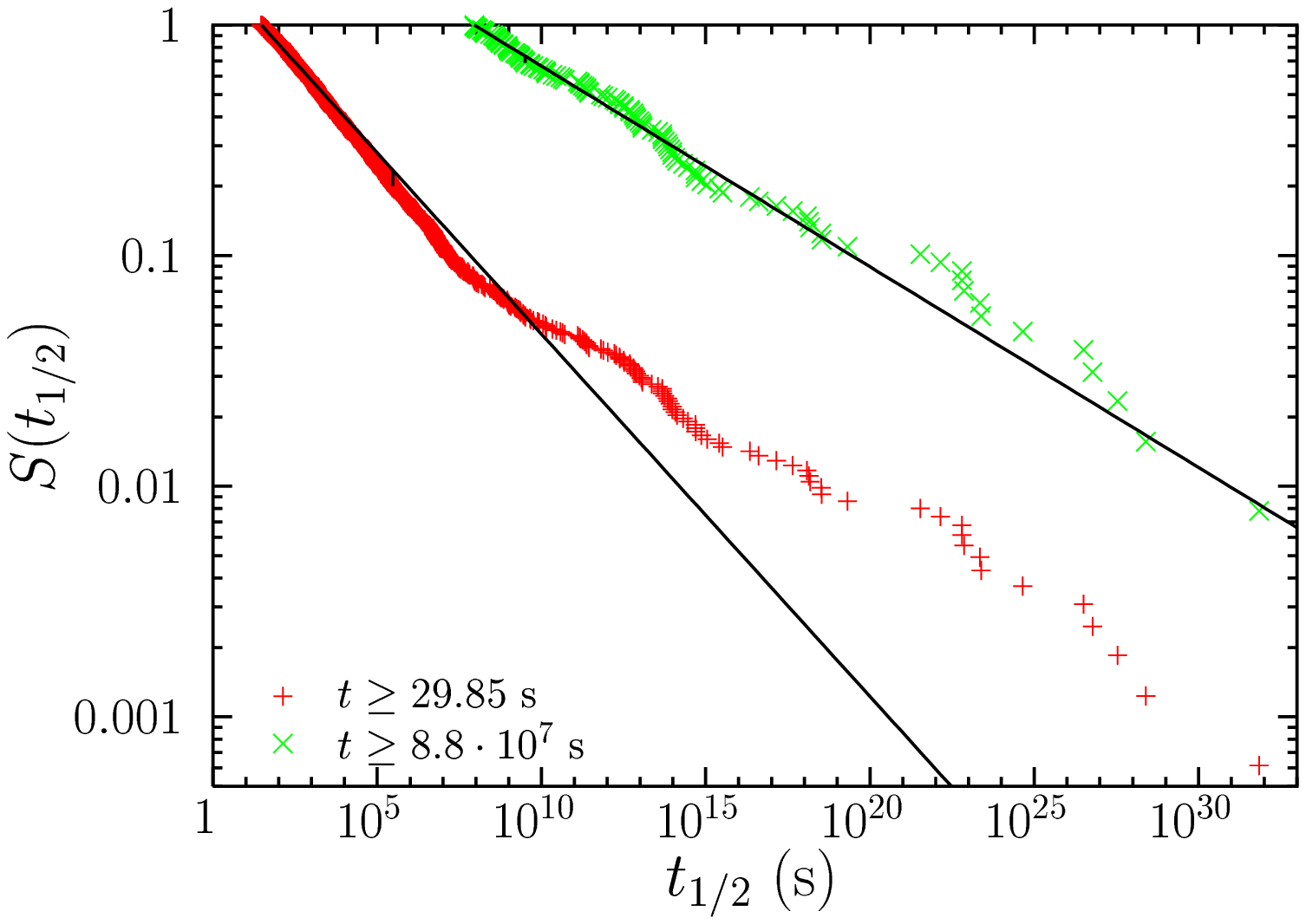}
\caption{
(Color online)
(a) Procedure for calculating the KS distance, exemplified with two data sets.
In each case, both the empirical and the fitted cumulative distribution 
functions are shown. The KS distance $d_m$ is the maximum vertical difference
between both curves, and the value of the half-life corresponding to this maximum
is shown as a vertical line. Two additional curves, given by $S(t) \pm d_m$, 
are also shown for clarity sake.
The first data set, on the left, is for the half-lives with $t \ge a = 29.85$ s,
corresponding to the Clauset {\it et al.}'s solution, 
which yields a bad fit, with $p=0$.
The second data set, on the right, is for $t \ge 8.8 \cdot 10^7 s$,
and although it yields a larger KS distance (0.052 versus 0.036),
leads to a much better power-law fit, 
as it is more clearly seen on 
(b) on a doubly logarithmic scale.
\label{fig_app}
}
\end{figure*}

As the estimated value of $\tau$ only depends on the geometric mean
of the data, any data set with the same $G_a$ will lead to the same
value of the exponent and the same uncertainty (if $N_a$ is fixed),
independently of whether the data come from a power law or 
from an alternative distribution.
Providing a goodness of the fit is necessary then.
For that purpose one can use the KS distance \cite{Press}, defined as, 
$$
d=\mbox{max}_{\forall t \ge a} |S(t)-S_{emp}(t)| ,
$$
i.e.,
as the maximum absolute difference between the empirical cumulative distribution function,
$S_{emp}(t)$, 
and the fitted cumulative distribution function, $S(t)$.
For simplicity, we identify the cumulative distribution with 
$S(t)=\int_t^\infty D(t)dt$
(that is, we use the complementary cumulative distribution, 
which does not affect the definition of the KS distance), and then
$$
S(t)=\left(\frac a t\right)^{\tau-1}, 
$$
while $S_{emp}(t)=n(t)/N_a$,
with $n(t)$ the number of data at or above $t$
(and not below $a$).
In this way, large values of $d$ denote bad fits, 
whereas small values correspond to good fits,
the boundary between large and small
will be made more precise below.
Figure \ref{fig_app} illustrates the computation 
of the KS distance for the nuclear half-lives with two special values of $a$,
taken from Table 1.

The key of the Clauset {\it et al.}'s recipe is to consider
the cutoff $a$ not as a fixed quantity, but as a parameter that needs to be 
estimated from data as well.
Then, the previous procedure (fitting of $\tau$ and calculation of KS distance) 
is repeated for all possible $a-$values, and the selected one corresponds
to the one that minimizes the KS distance,
i.e., $d_m =\min_{\forall a} d$.
This leads automatically to one value of $a$ and $\tau$.

In order to quantify the goodness of fit we need to compare $d_m$ with the results
for true power-law distributed data.
We simulate synthetic data sets, power law distributed for $t\ge a$, 
using 
$$
t=\frac{a}{(1-u)^{1/(\tau-1)}},
$$
with probability $N_a/N$ and
$u$ a uniform random number between 0 and 1,
and bootstraping the empirical data set for $t< a$, with probability $N_a/N$
(and $N$ is the total number of data, $\forall t$).
Then, we
apply exactly the same ML estimation of the exponent
and calculation of the KS distance to each synthetic data set.
We stress that the KS distance is computed between the
simulated distribution and its ML fit (not the fit of the empirical distribution,
which provides the parameters for the simulation).
In this way we end with a distribution of values of $d_m$,
which allows one to compute the $p-$value of the fit, as the ratio between 
the number of simulations with $d_m$ above the empirical one
and the total number of simulations.

If we generalize the method to an upper truncated power-law distribution, 
$$
D(t) = \frac{\tau-1}{1-r^{\tau-1}} \, \frac{a^{\tau-1}} {t^\tau}
$$
defined in $[a,b)$, with $r=a/b$, then the previous formulas need to be replaced by
$$
\frac 1 {\tau-1} + \frac{r^{\tau-1}\ln r}{1-r^{\tau-1}}
-\ln\frac {G_{ab}} a =0,
$$
$$
\sqrt{N_{ab}}\, \varepsilon=\left[\frac 1 {(\tau-1)^2} - \frac{r^{\tau-1}\ln^2 r}
{(1-r^{\tau-1})^2}\right]^{-1/2},
$$
$$
S(t)=\frac 1 {1-r^{\tau-1}} \left[ \left(\frac a t\right)^{\tau-1} - r^{\tau-1}  \right], 
$$
$$
t=\frac{a}{[1-(1-r^{\tau-1})u]^{1/(\tau-1)}},
$$
for the ML estimation of $\tau$, its asymptotic standard deviation (taken from Ref. \cite{Aban}),
the complementary cumulative distribution, 
and the simulated values of the variable in the power-law region,
$a \le t < b$ (taken with probability $N_{ab}/N$),
respectively.

\newpage



\newpage

\begin{widetext}
\begin{table}
\begin{tabular}{|l|cccrcccc|}\hline
condition & $a$ (s) & $b$ (s) & $b/a$ & $N_{ab}$ & c & $\tau \pm \varepsilon$ & $d_m$ & $p$ (\%) \\  
\hline 
 $b=\infty$  &       29.85 &   $\infty$ & $\infty$ 
    &   1624 &   0.145 &        1.16$\pm$0.01 &        0.036  & 
 0 \\
 $b=\infty$, $a>10^3$ s &   8.8$\cdot 10^{7\phantom{-}}$ &   $\infty$ & $\infty$ 
   &     128 &        0.018 &        1.09$\pm$0.01 &        0.052  & 
      33 $\pm $       5\\
 $b=\infty$, $a>10^4$ s &   8.8$\cdot 10^{7\phantom{-}}$ &   $\infty$ & $\infty$ 
   &     128 &        0.018 &        1.09$\pm$0.01 &        0.052  & 
      37 $\pm $       5\\
 $b=\infty$, $a>10^5$ s &   8.8$\cdot 10^{7\phantom{-}}$ &   $\infty$ & $\infty$ 
   &     128 &        0.018 &        1.09$\pm$0.01 &        0.052  & 
      37 $\pm $       5\\
\hline
 $a>1$ s &      126 &  7.9$ \cdot 10^{6\phantom{0}}$ & 6.3$\cdot 10^{4\phantom{0}}$
&       1138 &        0.185 &        1.18$\pm$0.01 &        0.012  & 
      88 $\pm $       3\\
$b/a >10^8$ &      126 &   1.3$ \cdot 10^{13}$ & 1.0 $ \cdot 10^{11}$ 
&       1277 &        0.198 &        1.19$\pm$0.01 &        0.015 & 
      39 $\pm $       5\\
\hline
$b<1$ s   &   5.0$\cdot 10^{-5}$ &   0.32 & 6.3$\cdot 10^{3\phantom{0}}$
&        460 &        0.085 &   0.65$\pm$0.02 &        0.023  & 
      40 $\pm $       5\\
$b<1$ s, $b/a > 10^4 $  &   1.3$\cdot 10^{-6}$  &   0.50  & 4.0$\cdot 10^{5\phantom{0}}$&
        559 &   0.079 &   0.67$\pm$0.02 &        0.024  & 
      32 $\pm $ 5 \\
\hline
\end{tabular}
\caption{
Results of several power-law fits and goodness-of-fit
for the half-lives of the 3002 radionuclides available.
The results for the non-upper-truncated power law ($b=\infty$) are calculated
with a resolution of 100 points per decade in $a$,
whereas for the upper-truncated power law we use 5 points per decade
in $a$ and $b$. 
$N_{ab}$ is the number of nuclides in the range $a \le t_{}< b$,
$c$ is the value of the proportionality constant of the power law
needed to fit the whole distribution in Fig. \ref{fig_density}
(when $t$ is measured in seconds), 
and $\varepsilon$ is one standard deviation in the distribution
of the maximum-likelihood estimation of the exponent $\tau$, see Ref. \cite{Aban}.
The $p-$value is obtained from 100 simulations, and
its uncertainty is calculated as in Ref. \cite{Corral_hurricanes}.
The value $p=0$ in the first row is maintained in 1000 simulations.
}
\end{table}
\end{widetext}

\begin{widetext}
\begin{table}
\begin{tabular}{|l|rrccrcccc|}\hline
case & $N$ & $a$ (s) & $b$ (s) & $b/a$ & $N_{ab}$ & c & $\tau \pm \varepsilon$ & $d_m$ & $p$ (\%) \\  
\hline
all &        3002 &      126 & 
 1.3$ \cdot 10^{13}$ &   1.0$ \cdot 10^{11}$ &        1277 & 
      0.198 &        1.19$\pm$0.01
 &        0.015 & 39$\pm$5 \\
\hline
ground &        2279 &      126 &   3.2$ \cdot 10^{13}$  & 2.5$ \cdot 10^{11}$ &
       1009 &   0.182 &        1.175$\pm$0.01
 &        0.020 &  18$\pm$4  \\
isomers &         723 &      200 &   7.9$ \cdot 10^{11}$  & 4.0$ \cdot 10^{9}$  &
        248 &   0.338 &        1.25$\pm$0.02 &        0.038  & 
      25$\pm $4 \\
\hline
$\alpha$ &         170 &      126 & 
 3.2$ \cdot 10^{25}$ &   2.5$ \cdot 10^{23}$ &          22 & 
 0.003 
&        1.00$\pm$0.01
 &        0.082 & 69$\pm$5 \\
$\beta^+$ &         607 &      501 & 
 3.2$ \cdot 10^{13}$ &   6.3$ \cdot 10^{10}$ &         281 & 
      0.833 &        1.29$\pm$0.02
&        0.036  & 30$\pm$5 \\
$\beta^-$ &         688 &      126 & 
 3.2$ \cdot 10^{17}$ &   2.5$ \cdot 10^{15}$ &         369 & 
 0.251 &        1.19$\pm$0.01
 &        0.028  & 37$\pm$5 \\
 EC &          82 &   \, 2$ \cdot 10^{5}$   
&   5.0$ \cdot 10^{13}$   & 2.5$ \cdot 10^{8\phantom{0}}$  & 
         62 &        2.289 &        1.22$\pm$0.03
 &        0.089  & 
       6 $\pm$2  \\
IT &         192 &        2 & 
 7.9$ \cdot 10^{9\phantom{0}}$ &   4.0$ \cdot 10^{9\phantom{0}}$ &         106 & 
 0.103 &        1.16$\pm$0.02
 &        0.053  & 42$\pm$5 \\
\hline
 \end{tabular}
 \caption{
Results of a power-law fit to the half-lives of nuclides 
in their ground or excited states (isomers)
and to the nuclides that follow a certain 
unique decay type. 
The fits are conditioned to $a> 1$ s and $b/a> 10^8$.
EC denotes orbital electron capture and IT, isomeric transition 
($\gamma$ ray or conversion electron emission from an exited state).
Other disintegration types yield very low numbers of nuclides, $N$.
Rest of symbols are as in the previous table.}
\end{table}
\end{widetext}


\end{document}